\documentclass[pra,twocolumn,superscriptaddress,showpacs,amsmath,amstex,amssymb,citeautoscript]{revtex4-1}

\usepackage[T1]{fontenc}
\usepackage[utf8]{inputenc}
\usepackage{lipsum}
\usepackage{amsmath}
\usepackage{amssymb}
\usepackage{bbm}
\usepackage{braket}
\usepackage{afterpage}
\usepackage{xcolor}
\usepackage{pifont}
\usepackage{stfloats}
\usepackage[mathscr]{euscript}
\usepackage[shortlabels]{enumitem}
\usepackage[justification=justified]{subcaption}
\usepackage{graphicx}
\allowdisplaybreaks
\usepackage{dsfont}
\usepackage{comment}
\usepackage[colorlinks=true]{hyperref}  
\hypersetup{
    bookmarks=true,         
    unicode=false,          
    pdftoolbar=true,        
    pdfmenubar=true,        
    pdffitwindow=false,     
    pdfstartview={FitH},    
    pdftitle={Adatom engineering magnetic order in superconductors},    
    pdfauthor={Pupim and Scheurer},     
    pdfsubject={},   
    pdfcreator={},   
    pdfproducer={}, 
    pdfkeywords={} {} {}, 
    pdfnewwindow=true,      
    colorlinks=true,       
    linkcolor=blue, 
    citecolor=blue,        
    filecolor=magenta,      
    urlcolor=blue           
}

\newcommand{\equref}[1]{Eq.~(\ref{#1})}

\newcommand{\figref}[1]{Fig.~\ref{#1}}
\newcommand{\refcite}[1]{Ref.~\onlinecite{#1}} 
\newcommand{\refscite}[1]{Refs.~\onlinecite{#1}}

\newcommand{\appref}[1]{Appendix~\ref{#1}}

\newcommand{\pdagger}{{\phantom{\dagger}}}

\renewcommand{\vec}[1]{\boldsymbol{#1}}

\definecolor{wrongultramarine}{rgb}{1,0.5,0}

\begin{document}

\title{Adatom engineering magnetic order in superconductors: \\ Applications to altermagnetic superconductivity}

\author{Lucas V.~Pupim}
\affiliation{Institute for Theoretical Physics III, University of Stuttgart, 70550 Stuttgart, Germany}

\author{Mathias S.~Scheurer}
\affiliation{Institute for Theoretical Physics III, University of Stuttgart, 70550 Stuttgart, Germany}

\begin{abstract}
We study theoretically how superlattices based on adatoms on surfaces of unconventional superconductors can be used to engineer novel pairing states that break time-reversal symmetry and exhibit non-trivial magnetic point symmetries. We illustrate this using a square-lattice Hubbard model with $d$-wave superconductivity and a subleading $s$-wave state as an example. An adatom superlattice with square-lattice symmetries is shown to stabilize an ``orbital-altermagnetic superconductor'', a state that exhibits loop current patterns and associated orbital magnetic moments, which preserve superlattice translations but are odd under four-fold rotations. This state is further characterized by a non-zero Berry curvature quadrupole moment and, upon including spin-orbit coupling, by an altermagnetic spin splitting of the bands and non-trivial spin textures in the superlattice unit cell, with zero net spin moment.
\end{abstract}

\maketitle

Time-reversal symmetry ($\mathcal{T}$) plays a special role for superconductivity \cite{AndersonPaper} as it guarantees the degeneracy of Kramers' partners which can thus form Cooper pairs at weak coupling. This is why pairing in the absence of $\mathcal{T}$ is rare in nature and why the coexistence of magnetism and superconductivity can give rise to a variety of interesting phenomenology, subject of intense current research \cite{SDE_new,daido_intrinsic_2022,he_phenomenological_2022,yuan_supercurrent_2022,Scammell_2022,OurTBG,2024arXiv240609505W,Han2024a,2024arXiv240701513S,2024arXiv240910712D,mazinNotesAltermagnetismSuperconductivity2022, 2024arXiv240317050B,2023arXiv230914812X,banerjeeAltermagneticSuperconductingDiode2024,jeschkeHighlyUnusualDoublystronglycorrelated2024,beenakker_phase-shifted_2023,brekke_two-dimensional_2023,chakraborty_zero-field_2023,ouassou_dc_2023,majorana1,majorana2,majorana3,andreev,giil2023superconductoraltermagnet,orientation_altermagnet,wei2023gapless}. Similarly, superconductors where the condensate itself spontaneously breaks $\mathcal{T}$ are actively being explored \cite{Ghosh_2021,wu2020superconductivity,ming2023evidence,TwistedCupratesExp} and yet there are only a few systems with clear experimental signatures. 


To stabilize the microscopic coexistence and probe the intricate interplay of superconductivity and magnetism in a controlled setting, here we propose to employ superlattices built by arranging non-magnetic adatoms on surfaces of materials; these can be assembled with current scanning probe techniques and have been shown to realize a variety of other interesting physics \cite{PhysRevResearch.2.043426,kempkes2019design,kempkes2019robust,slot2017experimental,doi:10.1021/acsnano.0c05747,PhysRevX.9.011009,drost2017topological,yahyavi2019generalized,zeng2016generalized,PhysRevB.103.235414}. The key ingredient in our work is a correlated substrate featuring unconventional superconductivity, where multiple competing superconducting channels naturally arise. Typically, details of the system, e.g., the chemical potential, determine the leading pairing state. Intuitively, these details can be altered locally by impurities or adatoms so that different superconducting states are favored in different spatial regions of a single sample, and, generically, a complex superposition of the order parameters minimizes the energy \cite{li2021superconductor,PhysRevB.105.014504,BrokenTComplPhases}, which in turn breaks $\mathcal{T}$. Employing this guiding principle, we show that appropriate adatom superlattices lead to complex magnetic order in the superlattice unit cell inside the superconducting phase. 

\begin{figure}
    \centering
    \includegraphics[width=\linewidth]{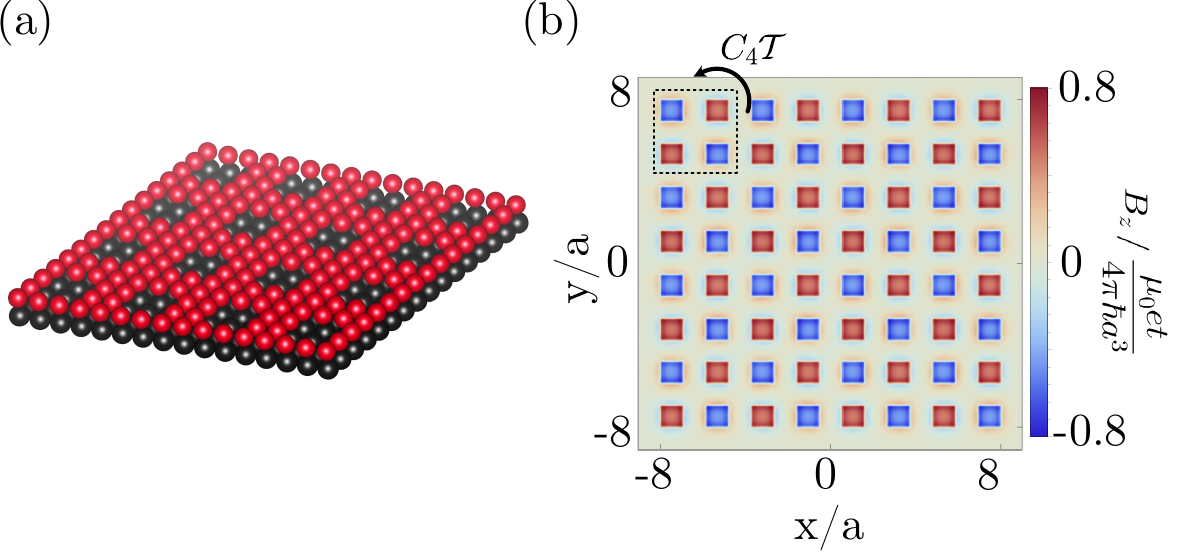}
    \caption{(a) Schematic of a superconductor surface (black) with a periodic arrangement of non-magnetic adatoms (red) on top. Given that the superconductor has competing orders, time-reversal symmetry can be spontaneously broken and generate loop currents. (b) Out-of-plane magnetic field component due to the loop currents. The dashed square shows the unit cell and we indicate the $C_{4z}\mathcal{T}$ symmetry, characterizing the orbital altermagnet. The chemical potential here is $\mu=-0.5$ and the magnetic field is calculated at $z=0.1 a$.}
    \label{fig:1}
\end{figure}

A currently very prominent form of magnetic order is altermagnetism \cite{yuan_giant_2020,smejkal_beyond_2022,smejkal_emerging_2022,ahn_antiferromagnetism_2019,bhowal_magnetic_2022,cuono_orbital-selective_2023,guo_spin-split_2023,maier_weak-coupling_2023,mazin_altermagnetism_2023,oganesyan_quantum_2001,sato_altermagnetic_2023,steward_dynamic_2023,turek_altermagnetism_2022,ouassou_dc_2023,mazin_prediction_21,hayami_momentum_19,smejkal_crystal_20,brekke_two-dimensional_2023,das_realizing_2023,fakhredine_interplay_2023,PhysRevLett.133.106701,leeb_spontaneous_2023,mazin_induced_2023,smejkal_chiral_2023,sun_spin_2023,gao_ai-accelerated_2023,beenakker_phase-shifted_2023,thermal_transport,RafaelsPaper,andreev,zhang2023finitemomentum,giil2023superconductoraltermagnet,majorana1,majorana2,majorana3,orientation_altermagnet,antonenko2024mirror,yu2024altermagnetism,wei2023gapless,mcclarty2023landau,banerjeeAltermagneticSuperconductingDiode2024,yershovFluctuationInducedPiezomagnetism2024,yershovFluctuationInducedPiezomagnetism2024,PhysRevB.108.L180401,2024arXiv240218629C,2024arXiv240910034J,2024arXiv240910712D,2024arXiv240701513S,2023arXiv230914812X,2024arXiv240317050B,2024arXiv240715836O,2024arXiv240215616R, zhaokondo2024,aoyama_piezomagnetic_2023,bai_efficient_2023,krempasky_altermagnetic_2023,lee_broken_2023,reimers_direct_2023,feng_anomalous_2022,bose_tilted_2022,2024arXiv240913504L, Ma2021May, ferrariAltermagnetismShastrySutherlandLattice2024,OurOrbitalAM,mazinNotesAltermagnetismSuperconductivity2022,jeschkeHighlyUnusualDoublystronglycorrelated2024,chakraborty_zero-field_2023}, which is characterized by local moments which average to zero over the lattice---not as a result of translational symmetry (like in an antiferromagnet) but due to (magnetic) point symmetries. For instance, a magnetic texture on the square lattice that is translationally invariant but odd under four-fold rotations $C_{4z}$ (and thus preserves $C_{4z}\mathcal{T}$) defines such an altermagnetic state \cite{smejkal_beyond_2022}. In particular, studying the impact of altermagnetism on superconductivity (assumed to coexist or to be induced via the proximity effect) has taken center stage in the field \cite{2024arXiv240701513S,2024arXiv240910712D,mazinNotesAltermagnetismSuperconductivity2022, 2024arXiv240317050B,2023arXiv230914812X,banerjeeAltermagneticSuperconductingDiode2024,jeschkeHighlyUnusualDoublystronglycorrelated2024,beenakker_phase-shifted_2023,brekke_two-dimensional_2023,chakraborty_zero-field_2023,ouassou_dc_2023,majorana1,majorana2,majorana3,andreev,giil2023superconductoraltermagnet,orientation_altermagnet,wei2023gapless}. 

Our work demonstrates an alternate route to the coexistence of superconductivity and altermagnetism by using adatoms to make the superconducting order parameter to become itself altermagnetic: we show that in the adatom superlattice in \figref{fig:1}(a), which enlarges the unit cell but still exhibits square-lattice symmetries, the superconductor spontaneously breaks $\mathcal{T}$ and $C_{4z}$ but preserves their product; since translations are simultaneously preserved, this defines what we call an ``altermagnetic superconductor’’. It exhibits local magnetic moments and fields, see \figref{fig:1}(b), coming from orbital currents, which average to zero due to $C_{4z}\mathcal{T}$. Adding spin-orbit coupling (SOC) leads to an altermagnetic spin texture in the unit cell.

\textit{Model}---Although the above-mentioned scheme of inducing complex magnetic order in an unconventional superconductor via superlattices of non-magnetic adatoms is more generally valid, here we illustrate it employing a square-lattice (lattice constant $a$) tight-binding model for concreteness. As used in \refcite{li2021superconductor} to describe superconductivity in the cuprates, we take a nearest-neighbor antiferromagnetic ($\mathcal{J}>0$) spin-spin interaction, $\mathcal{J}\sum_{\langle i,j\rangle} \vec{S}_i\cdot \vec{S}_j$, where $\vec{S}_i = \sum_{\sigma,\sigma'} c^\dagger_{i,\sigma} \vec{\sigma}_{\sigma,\sigma'} c^\pdagger_{i,\sigma'}$ and $c_{i,\sigma}$ denotes the annihilation operator of an electron on site $i$ and of spin $\sigma$. 
Decoupling the interaction in the spin-singlet channel, we obtain the following mean-field Hamiltonian 
\begin{align}\begin{split}
    H =&\sum\limits_{i,\sigma} \left(W_i -\mu  \right) \,c^\dagger_{i,\sigma} c^\pdagger_{i,\sigma} - \sum\limits_{i,j, \sigma} (t_{i-j} + \delta t_{ij})  c^\dagger_{i,\sigma} c^\pdagger_{j,\sigma}  \\
    &+ \left[\sum\limits_{\langle i,j \rangle, \sigma} \Delta_{ij} \left(c^\pdagger_{i,\uparrow}c^\pdagger_{j,\downarrow} - c^\pdagger_{i,\downarrow}c^\pdagger_{j,\uparrow} \right) + \text{H.c.}\right]. \label{eq:Hamiltonian_real}
\end{split}\end{align}
Here we already allow for adatoms, which break translational symmetry so that in general both the onsite potential and the hopping amplitudes are corrected by spatially modulated contributions, denoted by $W_i$  and $\delta t_{ij}$, respectively. As for the bare hoppings, $t_{i-j}$, we restrict the model to nearest $t$ (and measure all energies in units of $t$, i.e., $t \equiv 1$) and next-nearest neighbor terms $t'$ with $t' < 0$ \cite{PhysRevLett.87.047003}. The superconducting order parameter $\Delta_{ij}$ is determined by the self-consistency equation $\Delta_{ij}= \mathcal{J}\braket{c_{i,\uparrow}c_{j,\downarrow} - c_{i,\downarrow}c_{j,\uparrow}}$; in our numerics, we find $\Delta_{ij}$ by iteration until convergence with precision of $10^{-10}$.

\begin{figure}[tb]
    \centering
    \includegraphics[width=\linewidth]{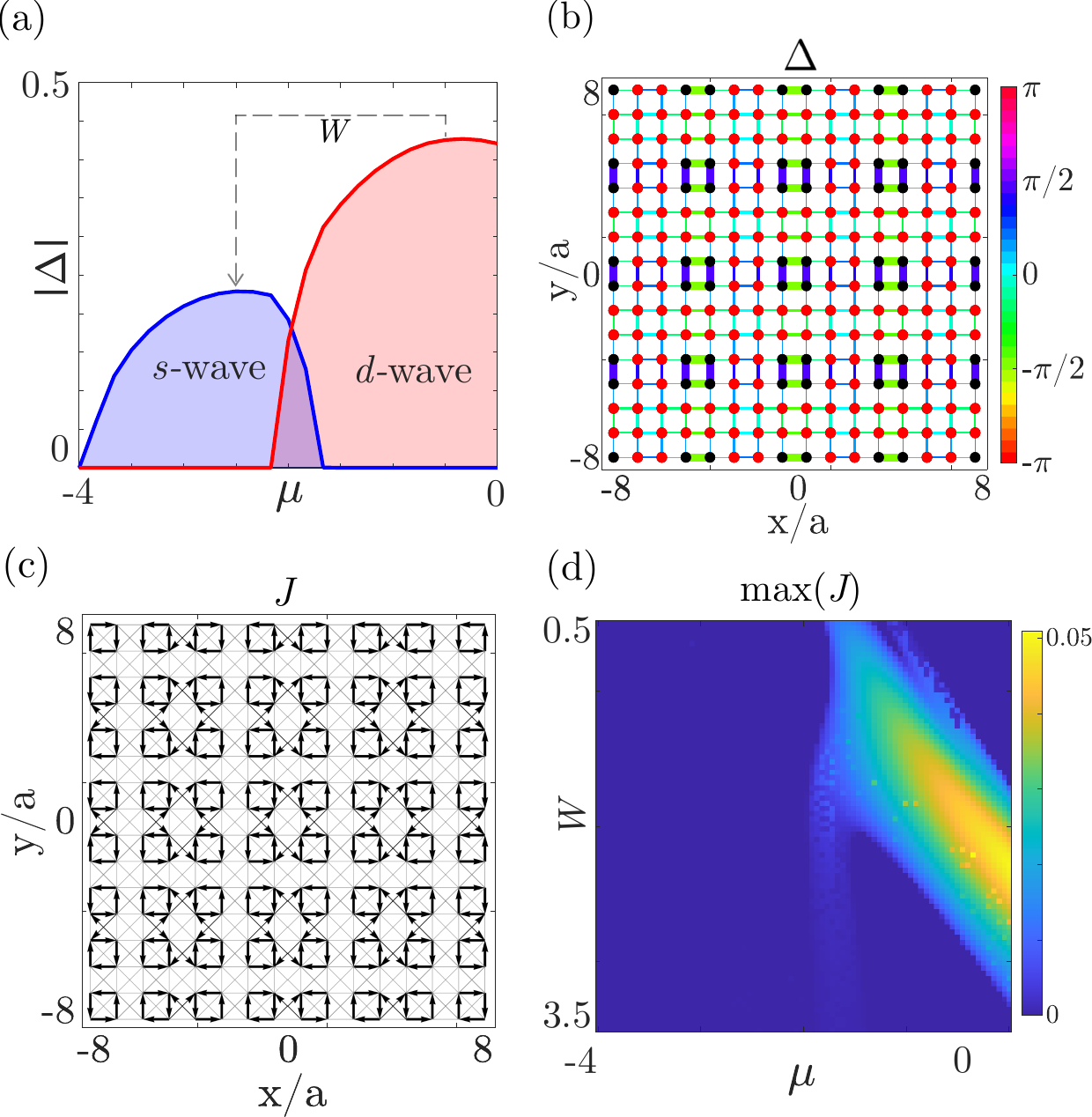}
    \caption{(a) Phase diagram without adatoms, featuring $d$-, extended $s$-wave, and $s\pm id$ superconductivity. The dashed line shows the two phases that are ``connected'' through a local adatom modulation $W$. (b)~Adatom configuration (red dots) on top of our superconducting lattice (black dots). The colors and width of the bonds represent the complex phase and magnitude of the resulting $\Delta_{ij}$. (c)~Respective currents, with only those larger than $5\%$ of the maximum current shown for clarity. (d) Maximal current for different $\mu$ and $W$ showing the robustness of the altermagnet phase.}
    \label{fig:2}
\end{figure}

\textit{Without adatoms---}We start by discussing the homogeneous limit, $W_i = \delta t_{ij} = 0$. Taking $\mathcal{J}=1.5$ and $t'=-0.1$, which we will assume throughout the main text, we either find a $d$-wave or an extended $s$-wave state as the leading superconductor, depending on the value of the chemical potential, see \figref{fig:2}(a). In both of those states, $\Delta_{ij}$ is only non-zero on neighboring sites and invariant under translation, with opposite signs (the same sign) for $d$-wave (extended $s$-wave) on bonds related by $C_{4z}$. Furthermore, for a small but finite range of $\mu$ [purple region in \figref{fig:2}(a))], these two orders coexist but with a relative phase of $\pi/2$, realizing an $s \pm i d$ superconductor \cite{PhysRevB.105.014504}. Intuitively, the non-trivial complex phase between the two orders can be thought as the system's tendency to maximize the excitation gap in this frustrated regime. Recalling the action of time-reversal $\mathcal{T}: \Delta^{\pdagger}_{ij}\rightarrow \Delta_{ij}^*$, we see that, out of these three superconductors, only the latter $s \pm i d$ state breaks $\mathcal{T}$ spontaneously.

\textit{Consequences of adatoms---}Now, with the role of $\mu$ on the nature of superconductivity clarified, we can foresee the consequences of local perturbations caused by adatoms or any other form of impurities \cite{PhysRevB.105.014504, li2021superconductor}. Referring to \appref{appendix} for the discussion of hopping modulations $\delta t_{ij}$, here we focus on spatial variations in the onsite potential $W_i$. It is no surprise that $W_i \neq \textit{const.}$ modulates our superconducting pairings $\Delta_{ij}$ spatially. However, because there are two competing order parameters ($s$- and $d$-wave), the spatial modulation can drive the superconducting energetics to break symmetries \textit{spontaneously}. All that is required is to set our system with $\mu$ lying in one of the phases and use adatoms capable of shifting the system (locally) to the other. 

We exemplify this idea through the dashed line in \figref{fig:2}(a): there, $\mu=-0.5$ so that, without any adatoms, our system is in the $d$-wave phase---the superconducting order parameter of the cuprate superconductors \cite{RevModPhys.72.969}. A modulation with $W=2$ tries to drive the sites with adatoms to the $s$-wave regime. Without coupling between the set of sites with and without adatoms, we would thus obtain regions with extended $s$- and $d$-wave superconductivity, respectively. Meanwhile, the coupling between the regions via the hopping terms and $\mathcal{J}$ determines the relative phases and, similar to the homogeneous $s + id$ phase, a configuration with non-trivial complex phases and broken time-reversal symmetry is favored due to the geometric frustration resulting from two order parameters. Moreover, the gradient of $\Delta$ (both in magnitude and its complex phase) can create loop currents, which can be viewed as Josephson currents \cite{li2021superconductor} and will be further studied below. Note that we are not using magnetic adatoms nor fine tuning of the system to the homogeneous $s \pm id$ phase. In fact, the $\mathcal{T}$-breaking superlattice superconductor also emerges for values of $t'$ where there is no $s \pm id$ phase in the phase diagram of \figref{fig:2}(a).

\textit{Adatom superlattice}---So far, the intuition we provided aligned with the conclusions in \refscite{PhysRevB.105.014504, li2021superconductor}, where random impurity distributions were studied. Instead, we henceforth explore periodic superlattices created by deliberate placement of adatoms, e.g., the red points in \figref{fig:2}(b) [and also in \figref{fig:1}(a)]. With this arrangement, the unit cell is enlarged by a factor of $4 \times 4$ but the associated superlattice has again the full square-lattice symmetry (with atomic translations promoted to superlattice translations). In \figref{fig:2}(b), we further see the resulting order parameter with its different phases and amplitudes varying inside the superlattice unit cell but explicitly invariant under superlattice translation. The non-trivial phases indicate broken $\mathcal{T}$ and the retained symmetries generating the magnetic point-symmetry group are: reflection through the axis ($\sigma_v$) and the composite symmetries $\sigma_d \mathcal{T}$ and $C_{4z} \mathcal{T}$ which represent, respectively, reflection through a diagonal and 4-fold rotation followed by time-reversal.

To demonstrate the non-trivial symmetry breaking through gauge-invariant observables, we calculate the currents
\begin{equation}
    J_{ij} = -i \frac{e \,t_{i-j}}{\hbar a^2} \sum\limits_\sigma \langle  c^\dagger_{i \sigma} c^\pdagger_{j \sigma} - \text{H.c.} \rangle
\end{equation}
on all bonds $(ij)$. The currents are found to be of the order of $\sim 0.04\frac{e t}{\hbar a^2}$ and their spatial pattern is shown in \figref{fig:2}(c). In agreement with the above symmetry analysis of the superconductor, they preserve superlattice translations and $\sigma_v$ but are odd under $C_{4v}$ and $\sigma_d$. 

Note that all orbital currents vanish, $J_{ij} = 0$, in the homogeneous $s \pm i d$ state as follows straightforwardly from the combination of magnetic point symmetries and invariance under translation by a single square-lattice site \cite{PhysRevLett.89.247003,PhysRevB.55.14554,PhysRevB.98.235126}. 
We also emphasize that the emergence of these orbital currents is a robust effect that does not require fine tuning. In \figref{fig:2}(d), we plot the maximum current on the lattice as a function of $\mu$ and $W$ for the same adatom superlattice as in \figref{fig:2}(b) and find an extended regime with $J_{ij} \neq 0$ and the same magnetic point symmetries as above.

\textit{Altermagnetic Superconductor}---Given the shape and arrangement of the orbital current loops, it is natural to associate them with Ising-like magnetic moments. To corroborate our intuition and compute the local magnetic field profile that can be measured in scanning magnetometry experiments, we consider each nearest- and next-nearest-neighbor link as a one-dimensional wire and apply the Biot-Savart law. As we can see in \figref{fig:1}(b), the resulting magnetic field exhibits well-separated regions of alternating sign arranged in a checkerboard-like pattern. This pattern again reveals the non-trivial magnetic point symmetries. Keeping the superlattice unit cell in mind, which is indicated as dashed black square in \figref{fig:1}(b), the regions of opposite sign of the magnetic field are not related by translational symmetry (like in an antiferromagnet) but by $C_{4z}$. As such, we refer to this superconductor as an \textit{orbital altermagnet}, where ``orbital'' refers to the fact that symmetries are broken in the orbital channel and the ground state is spin rotation invariant. The ``altermagnetic spin liquid'', recently discussed in \refcite{OurOrbitalAM}, is conceptually related since it also exhibits altermagnetic orbital currents while being spin-rotation invariant; however, the physical realizations---quantum disordering a non-coplanar magnet vs.~spatial frustration in a superconducting superlattice---differ significantly.

We note that these results are not restricted to the specific superlattice in \figref{fig:2}(b). It is also possible to observe this altermagnetic phase with different (bigger) superlattice unit cell sizes, which could help to probe the magnetic moments through magnetometry techniques, since it would require less spatial resolution. Deforming the square superlattice into a rectangular one is also possible, although now the magnetic point-symmetry group is reduced to $C_{2v}$; the residual $\sigma_v$ reflection symmetry is still sufficient to guarantee the vanishing of the total magnetic moment per unit cell. In addition, modulating the hopping, $\delta t_{ij} \neq 0$ in \equref{eq:Hamiltonian_real}, instead of the chemical potential, or starting with a different leading pairing state  lead to similar results (see Appendix).

\begin{figure}[tb!]
  \centering
  \includegraphics[width= 0.48\textwidth]{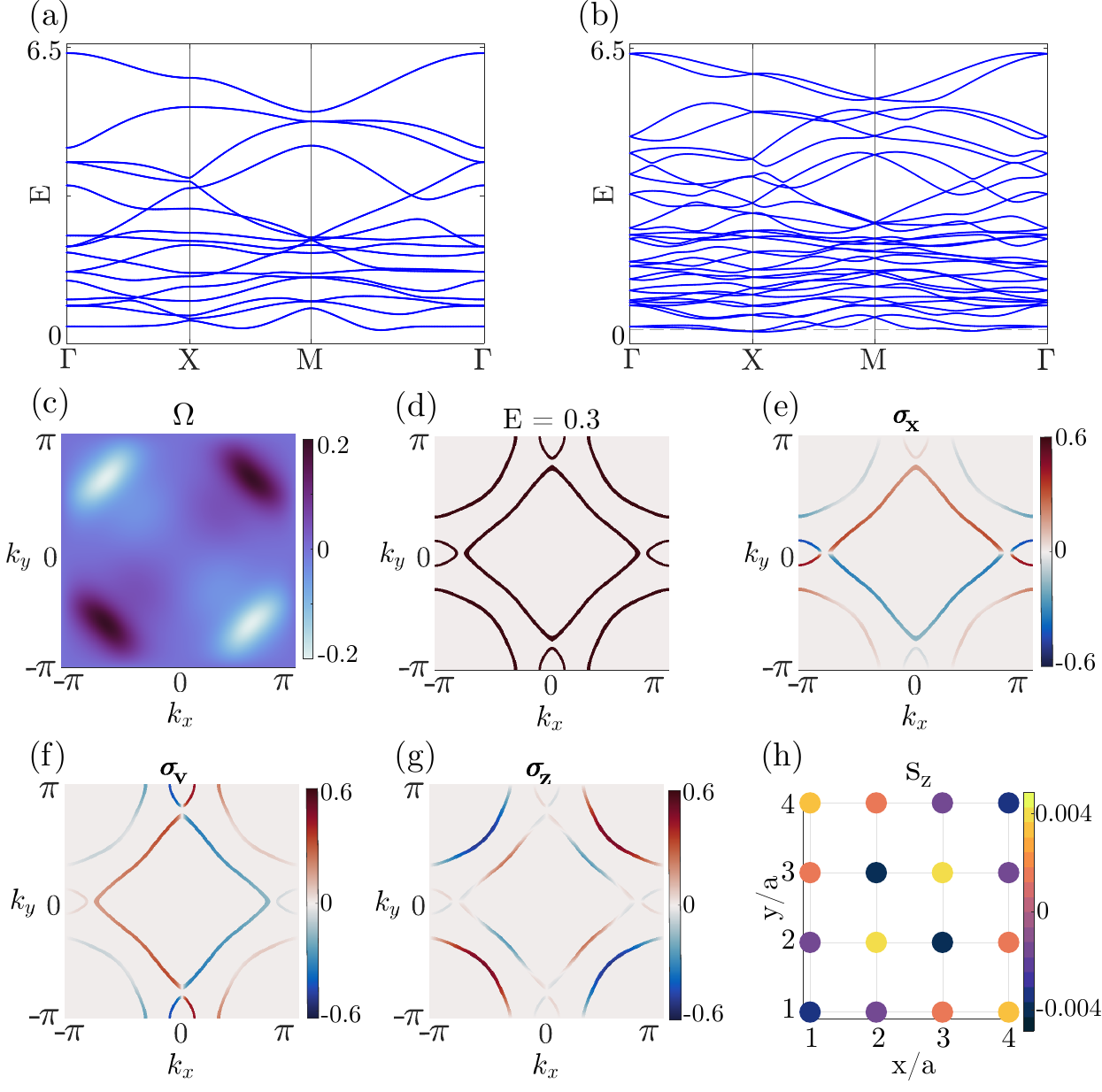}
\caption{Band structure of the altermagnetic superconductor (a) without and (b) with SOC ($\lambda = 0.2$). (c)~Composite Berry curvature of the occupied bands for $\lambda=0$. (d) Energy cut at $E=0.6$ from (b) and the respective spin texture (e-g) for the indicated spin components, which inherit the symmetries of the orbital altermagnet via $\lambda \neq 0$. (h) Local spins in the unit cell. They also obey the orbital altermagnet symmetries. The net spin in the unit cell vanishes, as expected for an altermagnet. We use $ \mu=-0.25$.}
\label{fig:3}
\end{figure}

To delve deeper into the superlattice superconductor, let us look into its band structure. To this end, we come back to the superlattice and order parameter of \figref{fig:1}(a) and \figref{fig:2}(b), where each superlattice unit cell contains $N=4\times4$ sites of the original square lattice. Each site $i$ within a given unit cell $\vec{R}$ is associated with a Nambu spinor $\psi_i(\vec{R}) =  [c^\pdagger_{i+\vec{R},\uparrow} \,\, c^\pdagger_{i+\vec{R},\downarrow} \,\, c_{i+\vec{R},\uparrow}^{\dagger} \,\, c_{i+\vec{R},\downarrow}^{\dagger} ]$, allowing us to subsequently describe superconductivity in the whole unit cell by $\Psi_{\vec{R}} = \left[ \psi_1(\vec{R}) \,\, \psi_2(\vec{R}) \,\, ... \,\, \psi_i(\vec{R}) \,\, ...\,\,\psi_N(\vec{R})  \right]$. We can now use this formulation to write
\begin{equation}
    H= \sum\limits_{\vec{R}} \big( \Psi^\dagger_{\vec{R}} h_0 \Psi^\pdagger_{\vec{R}} 
    +\sum\limits_{\nu=x,y} \left[ \Psi^\dagger_{\vec{R}} T_\nu \Psi^\pdagger_{\vec{R}+ \hat{\vec{e}}_{\nu}} + \text{H.c.} \right]  \big), \label{eq:Hamiltonian_SL_real}
\end{equation}
where all the terms from  \equref{eq:Hamiltonian_real} that represent processes within the unit cell are contained in $h_0$ and $T_\nu$ represents the hopping and pairing from one superlattice unit cell $\vec{R}$ to an adjacent cell at $\vec{R} + \hat{\vec{e}}_\nu$ with $\nu=x,y$. After Fourier transformation with respect to $\vec{R}$, $\Psi_{\vec{R}}=\sum_{\vec{k}} e^{-i\vec{k}\vec{R}}\tilde{\Psi}_{\vec{k}}$, and diagonalization of the $N \times N$ Bogoliubov-de Gennes Hamiltonian, we obtain the superconducting bands in \figref{fig:3}(a). Since the spectrum is particle-hole symmetric, we just plot positive energy bands for clarity. Only at $\Gamma$ and $M$ (which retain the full magnetic point group of the superconductor) are enough symmetries present to guarantee two-dimensional representations leading to degeneracies of bands; these are lifted away from these two high-symmetry points.

Since there is no $\vec{k}$-space-local anti-unitary symmetry (like $C_{2z}\mathcal{T}$, which is broken spontaneously by the superconductor), the Berry curvature $\Omega(\vec{k})$ can be non-zero. However, $C_4\mathcal{T}$ enforces $\Omega(\vec{k}) = -\Omega(C_{4z}\vec{k})$ such that both the Chern number (monopole) and the Berry curvature dipole moment have to vanish. The first non-zero moment is thus the Berry curvature quadrupole, which is confirmed by our explicit calculation \cite{fukui2005chern,Zhao:20} of the composite Berry curvature $\Omega$ for the ground state, shown in \figref{fig:3}(c). Our result further obeys the additional constraints $\Omega(\vec{k}) = \Omega(\sigma_d\vec{k}) = -\Omega(\sigma_v\vec{k})$ coming from the remaining magnetic point symmetries. While $C_4$ and $\sigma_v$ would have to be broken (e.g., by nematic order) to obtain a finite Chern number and, thus, finite thermal Hall effect, higher moments of the Berry curvature also bring observable signatures in non-linear transport, even with a trivial Chern number \cite{korrapati2024nonlinear,PhysRevX.14.021046,sodemann2015quantum,PhysRevLett.133.106701,ezawa2024intrinsicnonlinearconductivityinduced}. For our case of a superconductor, the non-zero Berry quadrupole moment can be probed via the non-linear thermal Hall effect and non-linear Nernst effect \cite{korrapati2024nonlinear}, which are observed under a temperature gradient.

\textit{Spin-orbit coupling}---We can make the spin channel inherit the altermagnetic structure, which was previously an entirely orbital effect, by adding SOC to our model. For concreteness, here we discuss a Rashba-like term \cite{bychkov1984properties} which reads as $i \lambda \left(c_{(x,y)}^\dagger \,\sigma_x \, c^\pdagger_{(x,y+1)} - c_{(x,y)}^\dagger\, \sigma_y \, c^\pdagger_{(x+1,y)} \right) + \text{H.c.}$ and respects all symmetries of the underlying square lattice and $\mathcal{T}$. Including the respective contributions in \equref{eq:Hamiltonian_SL_real}, we obtain the corresponding Bogoliubov bands shown in \figref{fig:3}(b). We can see that SOC splits the previously spin-degenerate bands everywhere but at the high-symmetry points. 

We can further see this splitting through an energy cut in \figref{fig:3}(d), and in (e) and (f) where we show the in-plane spin polarizations at the given energy. Note that the presence of $C_{2z}$ symmetry implies that the in-plane spin polarizations are always even under $\mathcal{T}$. They, thus, represent the Rashba SOC and do not contain any altermagnetic contribution. 
This is different for the out-of-plane spin component, which has to be odd under $\mathcal{T}$ and, hence, captures the altermagnetic part of the spin splitting, inherited from the superconductor. Indeed, we can see in \figref{fig:3}(g) that this component exhibits the sign changes under $C_{4z}$ with nodes pinned by $\sigma_v$ along the momentum axes, characteristic of a square-lattice altermagnet. Hence, by adding an in-plane SOC to our (out-of-plane) orbital altermagnet, we obtain an out-of-plane spin altermagnet. 
This result is true for all bands, but the degree of out-of-plane polarization can vary (roughly) from $10^{-4}$ to $0.6$ depending on the energy cut.

Interestingly, a non-zero local spin expectation value, $\braket{\vec{S}_{i}^z} \neq 0$, at a given site $i$, is also inherited from the orbital altermagnetic superconductor when including SOC, $\lambda\neq 0$. In \figref{fig:3}(h), we show the intra-unit cell behavior of the out-of-plane component $\braket{\vec{S}_{i}^z}$, which exhibits a non-trivial spatial texture that, by virtue of being odd under $C_{4z}$, has vanishing total magnetization.

\textit{Conclusion and outlook}---In this work, we demonstrated how superlattices of non-magnetic adatoms can be used to design non-trivial symmetry breaking in superconductors with multiple competing order parameters. In our concrete square-lattice computation, we showed that this can stabilize a novel altermagnetic superconductor with interesting and experimentally accessible properties. More generally, this approach can be applied to a variety of different superconducting systems which, apart from designing other symmetry-reduced superconductors of interest, can also be used to look ``below the surface'' of the leading superconducting state and uncover competing phases. At the same time, it can provide a tool for investigating the origin of signatures of broken time-reversal symmetry in certain superconducting compounds \cite{Ghosh_2021}.

\begin{acknowledgments}
The authors acknowledge funding by the European Union (ERC-2021-STG, Project 101040651---SuperCorr). Views and opinions expressed are however those of the authors only and do not necessarily reflect those of the European Union or the European Research Council Executive Agency. Neither the European Union nor the granting authority can be held responsible for them. M.S.S.~further acknowledges support by grant NSF PHY-2309135 for his stay at the Kavli Institute for Theoretical Physics (KITP) where part of the research was carried out. The authors thank J.~Sobral and S.~Banerjee for fruitful discussions.
\end{acknowledgments}

\bibliography{draft_Refs}

\onecolumngrid

 \begin{appendix}

 \section{Different adatom configurations}
\label{appendix}
\begin{figure}[h]
    \centering
    \includegraphics[scale=0.8]{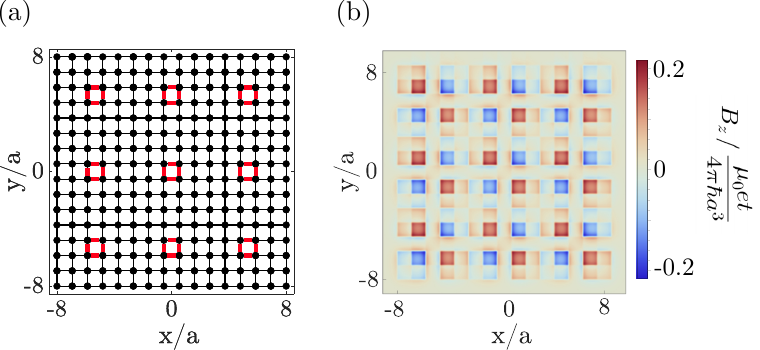}
    \caption{(a) Adatom placement affecting the hopping $t_{ij}$ on the red links. Here $\delta t_{ij} = -0.3$ instead of $t_{ij}=1$, which is valid for the black links. (b) Out-of-plane magnetic field component at $z=0.1 a$. Here $\mu = 1.8$. }
    \label{fig_app:1}
\end{figure}

In this appendix, we demonstrate the robustness of our conclusions in the main text by varying the form of the superlattice modulation.
The first configuration that we explore here can be thought of as placing the adatoms between two sites. In this situation, we assume that $W=0$ but $\delta t_{ij} \neq 0$ for $ij$ forming a square superlattice. We show in \figref{fig_app:1}(a) a possible arrangement of adatoms (red links) to realize this superlattice. In \figref{fig_app:1}(b) we can already see the resulting out-of-plane magnetic field component and that, again, the orbital altermagnet emerges but with a different periodicity (i.e., different unit cell size) compared to the examples in the main text. Here, however, there is a major difference from the case with adatoms on top of sites: the order parameter modulation is ``indirect'', i.e., we are not directly tuning the chemical potential. The drawback of this is that we need to keep $\mu$ close to the region with $s +id$ phase.

\begin{figure}[tbh]
    \centering
    \includegraphics[scale=0.8]{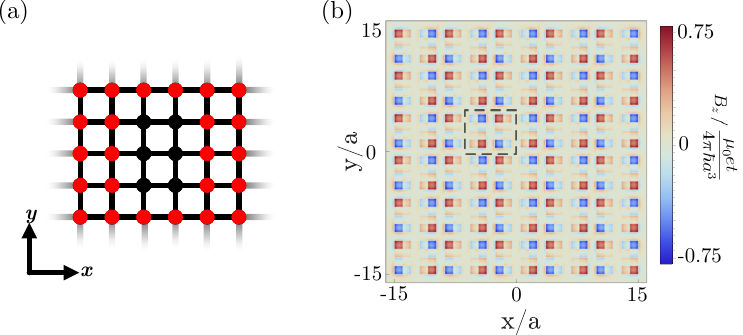}
    \caption{(a) Unit cell for a rectangular adatom superlattice. The red points show where the adatoms are placed. (b) Out-of-plane magnetic field component at $z=0.1 a$. Here $\mu=-0.5$ and $W=2$. The dashed line shows an unit cell.}
    \label{fig_app:2}
\end{figure}

As mentioned in the main text, we can also use a rectangular superlattice to modulate the superconductivity and still obtain an orbital altermagnet. Now, however, $C_{4z} \mathcal{T}$ and $\sigma_v$ are no longer present. In \figref{fig_app:2}(a) we illustrate the unit cell for this type of superlattice, while in (b) we show the respective out-of-plane component of the magnetic field.

\section{Starting from homogeneous $s$-wave }

Alternatively to what we show in the main text, one could start from a homogeneous system with extended $s$-wave superconductivity and use an adatom superlattice to modulate $\Delta$ to have a local preference for a $d$-wave symmetric order parameter. There are two ways to modify the system considered in the main text to achieve this modulation: one is to consider electron doped systems ($\mu>0$) and the second way is to use $W<0$ for hole doped systems. Here we use the first route to illustrate this alternative method.

\begin{figure}[h]
    \centering
    \includegraphics[scale=0.4]{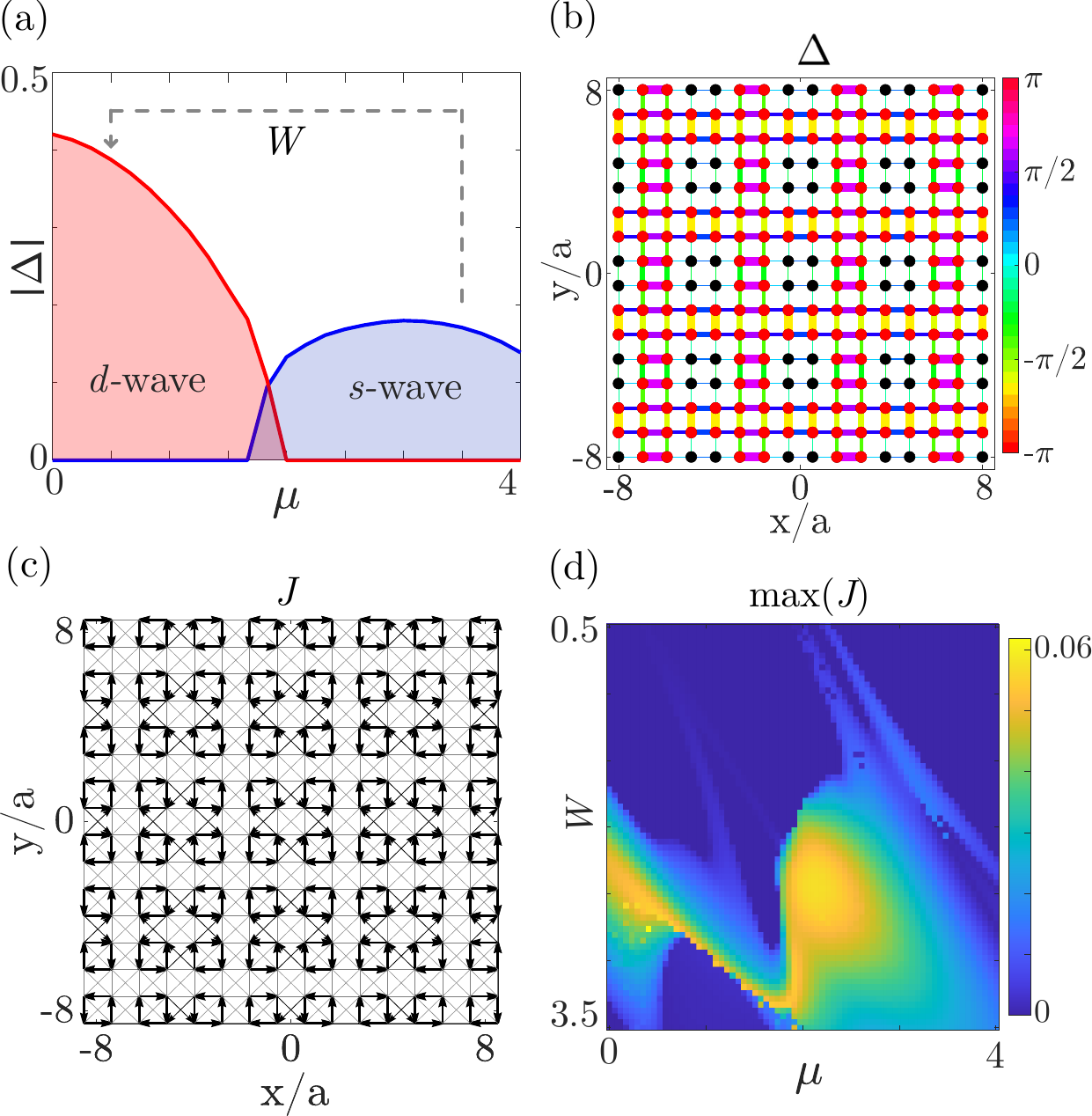}
    \caption{(a) Phase diagram without adatoms for electron doping, featuring $d$-, extended $s$-wave, and $s\pm id$ superconductivity. The dashed line shows the two phases that are ``connected'' through a local adatom modulation $W$. (b)~Adatom configuration (red dots) on top of our superconducting lattice (black dots). The colors and width of the bonds represent the complex phase and magnitude of the resulting $\Delta_{ij}$. (c)~Respective currents, with only those larger than $5\%$ of the maximum current shown for clarity. (d) Maximal current for different $\mu$ and $W$ showing the robustness of the altermagnet phase.}
    \label{fig_app:3}
\end{figure}

Similarly to the main text \figref{fig:2}, in \figref{fig_app:3}(a) we show the phase diagram without any adatoms. In (b) we observe the resulting $\Delta_{ij}$ for $\mu=3.5$ and notice that it has  the ``dual'' behavior from \figref{fig:2}(b), i.e., now $\Delta_{ij}$ is stronger in the links between sites with adatoms rather than the link between sites without adatoms. In (c), we show the respective current profile which is the same as the one in \figref{fig:2}(c). For \figref{fig_app:3}(d) we observe the biggest difference when comparing to the corresponding plot in \figref{fig:2}(d). Now the presence of loop currents is centered around the $s$-wave region as anticipated. We can also see that for the $d$-wave region there are also currents. This happens because in that region $W$ now connects the $d$-wave for $\mu>0$ with the $s$-wave region for $\mu<0$.Whenever there are orbital currents, the superconductor and currents exhibit the same altermagnetic symmetries.

 \end{appendix}

\end{document}